\documentclass{mytemplate}

\usepackage{url}
\usepackage{booktabs}  
\usepackage{amsfonts} 
\usepackage{microtype} 
\usepackage{amsfonts}
\usepackage{lmodern}
\usepackage{longtable}
\usepackage{setspace}
\usepackage{indentfirst}
\usepackage{courier}
\usepackage{listings}  
\usepackage{titletoc}
\usepackage{multirow}
\usepackage{listings}

\usepackage{algorithm}
\usepackage{algpseudocode}

\usepackage{enumitem} 
\usepackage{graphicx} 
\usepackage{amsmath} 
\usepackage{amsfonts, amsmath, amssymb, amsthm, thmtools}
\usepackage{multicol}
\usepackage{caption}
\usepackage{subcaption}
\usepackage{booktabs}
\usepackage{wrapfig}
\usepackage{placeins} 

\newcommand{\paper}[0]{paper}
\newcommand{\work}[0]{work}

\newcommand{\Article}[0]{Article}

\usepackage{listings}
\usepackage[natbib = true,
    backend=biber,
    isbn=true,
    url=false,
    doi=true,
    eprint=true,
    style=ieee
    ]{biblatex}
\usepackage{enumitem}
\setlist{nosep}
\setlist{itemsep=1pt, topsep=3pt}
\title{Monte Carlo simulation studies on Python using the sstudy package with SQL databases as storage}

\author[1]{Marco H A Inácio}
       
\affil[1]{University of São Paulo, Brazil;
       Federal University of São Carlos, Brazil and
       Budapest University of Technology and Economics, Hungary. Email: \url{m@marcoinacio.com}}

\begin{document}

\maketitle

\begin{abstract}
Performance assessment is a key issue in the process of proposing new machine learning/statistical estimators.
A possible method to complete such a task is by using Monte Carlo simulation studies, which can be defined as the procedure of estimating and comparing properties (such as predictive power) of estimators (and other statistics) by averaging over many replications given a true distribution; i.e.: generating a dataset, fitting the estimator, calculating and storing the predictive power, and then repeating the procedure many times and finally averaging over the stored predictive powers. 
Given that, in this paper, we present \textit{sstudy}: a Python package designed to simplify the preparation of simulation studies using SQL database engines as the storage system; more specifically, we present its basic features, usage examples and references to the its documentation.
We also present a short statistical description of the simulation study procedure with a simplified explanation of what is being estimated by it, as well as some examples of applications.

\end{abstract}

\section{Introduction}
One important aspect of proposing new machine learning/statistical estimators and methods is the performance test phrase.
A possible way to access such performance is by Monte Carlo simulation studies, which can be defined as the procedure of estimating and comparing properties (such as predictive power) of estimators (and other statistics) by averaging over many replications given a true distribution;
i.e.: generating a dataset, then fitting the estimator, calculating and storing the predictive power, and then repeating the procedure many times and finally averaging over the predictive powers across repetitions.

In this \paper, we present a Python\citep{python} package called \textit{sstudy} which is designed to simplify their preparation and execution using SQL database engines. We also present a short statistical description of the simulation study procedure, as well as some examples of application using the package.

\subsection{Related work}
While applications of simulation studies are abundant in literature (with a simple search of ``Monte Carlo simulation study'' on \textit{Google Scholar} yielding up to a thousand results), there is also a considerable literature regarding the discussion and meta analysis of Monte Carlo simulation studies themselves. In particular, \citet{Morris2019} presents a Biostatistics tutorial on the rationale behind using simulation studies, while providing guidance
for their design, execution, analysis, reporting and presentation, with special focus on method evaluation.

\citet{Burton2006} discuss, in the context of medical Statistics, issues to consider when designing a simulation study; in particular, by exposing how the study will be performed, analysed and reported in details.
Moreover, design decisions are discussed such as the procedures for generating datasets and the number of simulations to be performed.
The authors also suggest a checklist of important design considerations.

\citet{Metcalfe2005} discuss the effect of varying the event generation process (data generating distribution) on simulation studies in the context of evaluation of Statistical methods for the analysis of recurrent events. Four distinct generating distributions (Poisson, mixed Poisson, autoregressive, and Weibull) are used to evaluate a set of distinct statistical estimators and their impact in the results is analysed. The authors conclude that the event generation process impact the quality of the estimator and that, therefore, multiple generation processes should be considered on a study.

\citet{Mundform2011} discuss the choice of the number of replications (simulations) to be done for simulation studies with regards to the quality of the simulation, in terms of, for example, type I
error rates, power and run time. 22 works in literature were
analysed and replicated in order to find the minimum number of simulations to be done in order to achieve stable results.
The authors concluded that in many cases fewer simulations than the original ones used in the works were needed to produce stable estimates of the results, and that, for all works, less than 10000 simulations were sufficient to achieve stable results.

\citet{Schaffer2007} also discuss the choice of the number of replications to be done on simulation studies, but specifically in the context of control chart analysis. They also conclude than less than 10000 simulations are sufficient to achieve the desired performance in their desired criteria, and moreover, that in many cases less than 5000 were also enough.

Finaly, \citet{mooney1997monte} provides an extensive book on the subject and presents, among other things, the logic behind Monte Carlo simulation studies, a set of five steps to implement them and discusses their use in social sciences.

\subsection{Terminology}
Given the mixed audience nature of this \paper, we use the following terms interchangeably:

\begin{itemize}
    \item Train model, fit model to data.
    \item Dataset, sample.
    \item Number of instances, sample size.
    \item Dataset generator, true distribution, data generating function.
    \item Loss, decision criteria.
\end{itemize}

\subsection{{\Article} organization}
The rest of this \paper is organized as follows: 
in Section \ref{sec:on_sim_studies}, we present a short introduction on simulation studies, with a short statistical notation of what is being estimated by them.
In Section \ref{sec:the_package}, we present a short introduction on \textit{sstudy} package, with part of the source code of an example of basic usage as well as presenting its features and their usage examples available in its documentation.
In Section \ref{sec:sim_examples}, we present some examples of applications of simulation studies using the \textit{sstudy} package with results and analysis. The source code for such examples is distributed together with the package.
Finally, section \ref{sec:article_conclusion} concludes the \paper.

\section{On simulation studies:}
\label{sec:on_sim_studies}
The process of a simulation study consists of varying some aspects of the data generating function, the estimating model and estimating its performance by averaging over distinct random seeds for the data generator;
i.e., estimating:
\begin{align*}
    E_{D \in \mathbb{D}_P}\left[ \mbox{loss}(M_k(D), D) \right]
\end{align*}
where
\begin{itemize}
    \item P are the parameters of the data generating function (e.g.: distribution parameters, number of instances, etc).
    \item k are model parameters (e.g.: whether you are using a linear regression, a lasso or a ridge, and if a lasso/ridge, what is its tuning parameter, etc).
\end{itemize}

So, in other words, a simulation study is a repetition of the following procedure many times followed by averaging over the results: generate a dataset $D$ from a ground truth distribution $\mathbb{D}_P$, train a model $M_k$ using this dataset and then evaluate the loss.

Note however, that in order to avoid overfitting, one must train and evaluate the loss on distinct partitions of the dataset $D$\footnote{Note that, in some cases, however it is not necessary to generate a test dataset because a dataset might not be necessary at all in the evaluation phrase; e.g.: if you want to compare the estimated parameter directly to the true model parameter.}. Algorithm \ref{alg:sim_study} summarizes the procedure.

\begin{algorithm}
 \caption{ \small Simulation study procedure}\label{alg:sim_study}
 \textbf{Input:} {\small  
 dataset generator $\mathbb{D}_P$,
 model $M_k$,
 number of desired simulations $n\_sim$.
 } \\
 \textbf{Output:} {\small  
 loss over simulations.
 } 
 \begin{algorithmic}[1]
  \For{$i \in \{1,\ldots,n\_sim\}$}
      \State Generate dataset $D_{\mbox{train}}$ and $D_{\mbox{test}}$ from $\mathbb{D}_P$.
      \State Train model $M_k$ using $D_{\mbox{train}}$: i.e. calculate $M_k\left(D_{\mbox{train}}\right)$.
      \State Evaluate $\mbox{loss}\left(M_k\left(D_{\mbox{train}}\right), D_{\mbox{test}}\right)$ and store it on $L_i$
  \EndFor
  \State Return the mean of $L$.
 \end{algorithmic}
\end{algorithm}

\section{The sstudy package}
\label{sec:the_package}

In this section, we present the basic usage of the package as well as some of its features.

\subsection{The package}
The \textit{sstudy} Python package is a thin layer designed to simplify the preparation and execution of simulation studies using SQL database engines. The package works by randomly selecting a simulation study configuration and checking if this configuration has already seem the user requested number of simulations; then, if that's the case, skip to another randomly selected simulation configuration, otherwise, do the simulation study and store the results on the SQL engine.

Given the independence of each simulation study and the atomicity of SQL engines\footnote{Atomicity: this means that either a transaction (e.g.: storing the simulation study results) will happen entirely, or it won't happen at all.}, this procedure can be parallelized (i.e.: one can span multiple simulation study processes), even over multiple machines. Moreover, if the procedure is stopped abruptly, simulation studies previously done will be not affected as they are already stored in SQL database.

Note that this process of ``checking'' (and then assigning the simulation configuration to be done) in general involves only a few (mili)seconds (provided the communication with the SQL server is fast enough), so that in the general terms, the bottleneck of execution is the simulation studies themselves.

\subsection{Storage engines}
The package is build around the \textit{peewee} Python package which as of the date of this \work, supports SQLite, PostgreSQL, MySQL and CockroachSQL.

For simple project running on a single machine, SQLite is the recommend storage system as it is a self contained SQL engine which is stores the dataset in a single user-defined file and does not require the installation of a SQL server system. Moreover, SQLite files can be opened and explored using GUI tools such as the DB browser SQLite.

For projects running on a multiple machines, PostgreSQL (or MySQL) is the recommend storage system as it is a well supported open source SQL engine, although it requires the a server installation (or renting a pre-installed server provided by a cloud services platform). CockroachDB on the other hand is a distributed SQL system which can be stored on a cluster on machines.

\subsection{Basic usage}

The recommended design of a experiment using the \textit{sstudy} package is by having it separated in 3 files:
\begin{itemize}
    \item A file for database structure where we declare the variables to be stored in the SQL database and their respective types (see Listing \ref{lst:database_structure}).
    \item A file for running the simulations where we declare the list of parameters to be simulated, as well as the simulation script itself (see Listing \ref{lst:simulations}).
    \item A file to explore/plot the results which can be exported directly into a \textit{pandas.DataFrame} (see Listing \ref{lst:pandas}).
\end{itemize}

\begin{lstlisting}[float,floatplacement=hbtp,caption=Part of the database structure file,label=lst:database_structure]
class Result(Model):
    # Data settings
    data_distribution = TextField()
    method = TextField()
    no_instances = DoubleField()

    # Results
    score = DoubleField()
    elapsed_time = DoubleField()
\end{lstlisting}

\begin{lstlisting}[float,floatplacement=hbtp,caption=Part of the simulation execution file,label=lst:simulations]
to_sample = dict(
    data_distribution = ["complete", "sparse"],
    no_instances = [100, 1000],
    method = ['ols', 'lasso'],
)

def func(
    data_distribution,
    no_instances,
    method,
    ):
    
    x = (no_instances + 10000, 10)
    x = stats.norm.rvs(0, 2, size=x)
    beta = stats.norm.rvs(0, 2, size=(10, 1))
    eps = (no_instances + 10000, 1)
    eps = stats.norm.rvs(0, 5, size=eps)
    if data_distribution == "complete":
        y = np.matmul(x, beta) + eps
    elif data_distribution == "sparse":
        y = np.matmul(x[:,:5], beta[:5]) + eps
    else:
        raise ValueError

    y_train = y[:no_instances]
    y_test = y[no_instances:]
    x_train = x[:no_instances]
    x_test = x[no_instances:]

    start_time = time.time()
    if method == 'ols':
        reg = LinearRegression()
    elif method == 'lasso':
        reg = Lasso(alpha=0.1)
    reg.fit(x_train, y_train)
    score = reg.score(x_test, y_test)
    elapsed_time = time.time() - start_time

    return dict(
        score = score,
        elapsed_time = elapsed_time,
    )

do_simulation_study(to_sample, func, db, Result,
max_count=no_simulations)
\end{lstlisting}

\begin{lstlisting}[float,floatplacement=hbtp,caption=Part of the simulation study results exploration file,label=lst:pandas]
import pandas as pd
...
df = pd.DataFrame(list(Result.select().dicts()))
df.groupby(['data_distribution', 'no_instances',
'method']).mean()
\end{lstlisting}

To see the complete source code of listings \ref{lst:database_structure}, \ref{lst:simulations} and \ref{lst:pandas}, see the examples/basic folder distributed together with the package, which is also available at: \url{https://gitlab.com/marcoinacio/sstudy/-/tree/master/examples/basic}.

\subsection{Main features and documented examples}
In the package documentation available at \url{https://sstudy.marcoinacio.com/}, we present the following features and examples:

\begin{itemize}
    \item Support to SQLite, PostgreSQL, MySQL and CockroachDB (and, at least in principle, any additional dataset supported by \textit{peewee}).
    \item Automatic randomization of executions.
    \item Optional filter of undesired simulation options. 
    \item Prevention of SQL server disconnect failures: waits for availability of the server again so that long simulation calculations are not lost.
    \item Automatic handling of binary data: whenever a dataset field is a \textit{BlobField}, invokes the ``binarizer'' \textit{pickle.dumps} automatically. This allows the user to store whole arrays or large class instances as results into the SQL database.
    \item Hints on exploring the results using \textit{pandas} package.
\end{itemize}

\section{Examples of applications}
\label{sec:sim_examples}

In the following subsections, we present a series of the example of usage of simulation studies using the \textit{sstudy} package, the source code of all example is available to download in the package \textit{examples} folder at \url{https://gitlab.com/marcoinacio/sstudy/-/tree/master/examples}. A Dockerfile is also available at \url{https://gitlab.com/marcoinacio/sstudy/-/blob/master/Dockerfile} in order to install the dependencies and run all examples on Docker.

\subsection{Simple regression}
Suppose that we want to compare the performance of ordinary least squares with the performance of a lasso with data being generated from a Gaussian linear regression: e.g.: each dataset contains $100$ instances $(X_1, X_2, ..., X_{100})$, with each instance arising independently from a
$Y|X \sim \mbox{Gaussian}(X \beta, \sigma)$.
$X \sim \mbox{Multivariate Gaussian}(0, 2 I)$.

In other to proceed with the evaluation, one must note first that there are many possibilities that we could setup here in order to test the estimators performance: we could compare the estimated values of $\mu = X \beta$ or compare directly the estimated $y$'s. Moreover, we can also choose from a wide range of loss criteria\footnote{Without loss of generality, you can also work with utility, score and other decision criteria.} like the mean squared error, mean absolute error, etc.

A second point to notice here is that one might be tempted to generate a single train dataset $D_{\mbox{train}}$ (i.e.: a sample $(X_1, X_2, ..., X_{30})$) and a single test dataset $D_{\mbox{test}}$ (i.e.: a sample $(X_1, X_2, ..., X_{m})$ as large as we want), fit the models (empirical mean and median) to data, and evaluate their mean squared error. In this case however, we would be affected by random chance and would be unable to conclude with certainty which model bests adjust to the data: maybe model A is better for this true distribution, but by chance it happened to obtain a bad fit this specific dataset that was generated.

If however, we try to solve this problem by means of increasing the train dataset size, then we fall into another problem: we would be concluding which model better fits a large sample size instead of perceiving their behaviour on smaller samples\footnote{Note that models that have terrible behaviour on small datasets, might get increasingly better as the sample size increases (i.e.: bad estimators might be consistent). An example would be the estimator $\sum_{i=1}^nx_i) / (n-10000)$ which is generally bad for small samples but equals to the empirical mean as $n$ approaches infinity.}, but in the real world, we do not have access to an infinite amount of data.

The solution given by a simulation study is to repeat such a procedure many times to the point that in the long run, we are able to distinguish which model is the best for this true distribution and this decision criteria even in the presence of small datasets.

In Table \ref{tab:res_ss}, we present the results for such a simulation experiment (in parenthesis we present the standard error of measurement of the simulation study, note that if you increase the number of simulations, the standard error will tend to zero by the law of large numbers).

\begin{table}[htbp]
 \ttfamily
 \centering
 \caption{Results for a simulation experiment using a Gaussian linear regression as dataset generator.}
 
\begin{tabular}{crlc}
\toprule
data  &  n. of & \multirow{2}{*}{method} & \multirow{2}{*}{score} \\
distribution & instances \\
\midrule
\multirow{6}{*}{complete}& \multirow{2}{*}{100}   & lasso  &  0.803 (0.035)  \\
       &       & ols &  0.838 (0.008)\\
       & \multirow{2}{*}{1000}  & lasso &  0.856 (0.011)\\
       &       & ols &  0.834 (0.010) \\
       & \multirow{2}{*}{10000} & lasso &  0.842 (0.016)  \\
       &       & ols &  0.825 (0.012) \\
\multirow{6}{*}{sparse} & \multirow{2}{*}{100}   & lasso &  0.608 (0.073) \\
       &       & ols &  0.660 (0.022) \\
       & \multirow{2}{*}{1000}  & lasso &  0.747 (0.049)  \\
       &       & ols &  0.702 (0.022) \\
       & \multirow{2}{*}{10000} & lasso &  0.688 (0.040)  \\
       &       & ols &  0.695 (0.021) \\
\bottomrule
\end{tabular}
 \label{tab:res_ss}
\end{table}

The source code for this experiment is available to download in the package \textit{examples/basic} folder at \url{https://gitlab.com/marcoinacio/sstudy/-/tree/master/examples/basic}.

\subsection{P-values of hypothesis tests}
Simulation studies can also be used to compare the hypothesis testing methods (see \citet{vaecompare} and \citet{1908.00105}, for instance). In this case, two important criteria arises: the uniformity of the test under the null hypothesis and the test power under the alternative hypothesis.

Given a dataset $D_{\mbox{train}}$ (i.e.: a sample $(X_1, X_2, ..., X_{n})$) with each instance coming independently from a $\mbox{Gaussian}(\mu=2, 1)$, we could, for instance, compare the tests type 1 error rate of method A and B under the null hypothesis $\mu=2$ and compare the test power of such methods under the alternative hypothesis $\mu = 3.5$.

Additionally, we could change the true distribution to something other than a Gaussian to verify how that affects the type I error and the test power.

In order to illustrate this, we present an simulation study comparing the performance of two sample comparison methods, i.e.: hypothesis tests that take two datasets as input and attempt to test the hypothesis of whether the two datasets arise from the same data generating function. We work with methods
Mann-Whitney rank test\citep{Mann47},
Kolmogorov-Smirnov\citep{Smirnov48} and
Welch's t-test\citep{WELCH1947} for datasets with 1000 and 2000 instances, with each instance being generated by a standard log-normal distribution. Moreover, under the alternative hypothesis, one the datasets has 0.1 added to all of instances after sampling from the log-normal distribution. 1000 simulations were performed for each configuration.

\begin{figure}[hbtp]
\includegraphics[width=0.75\columnwidth]{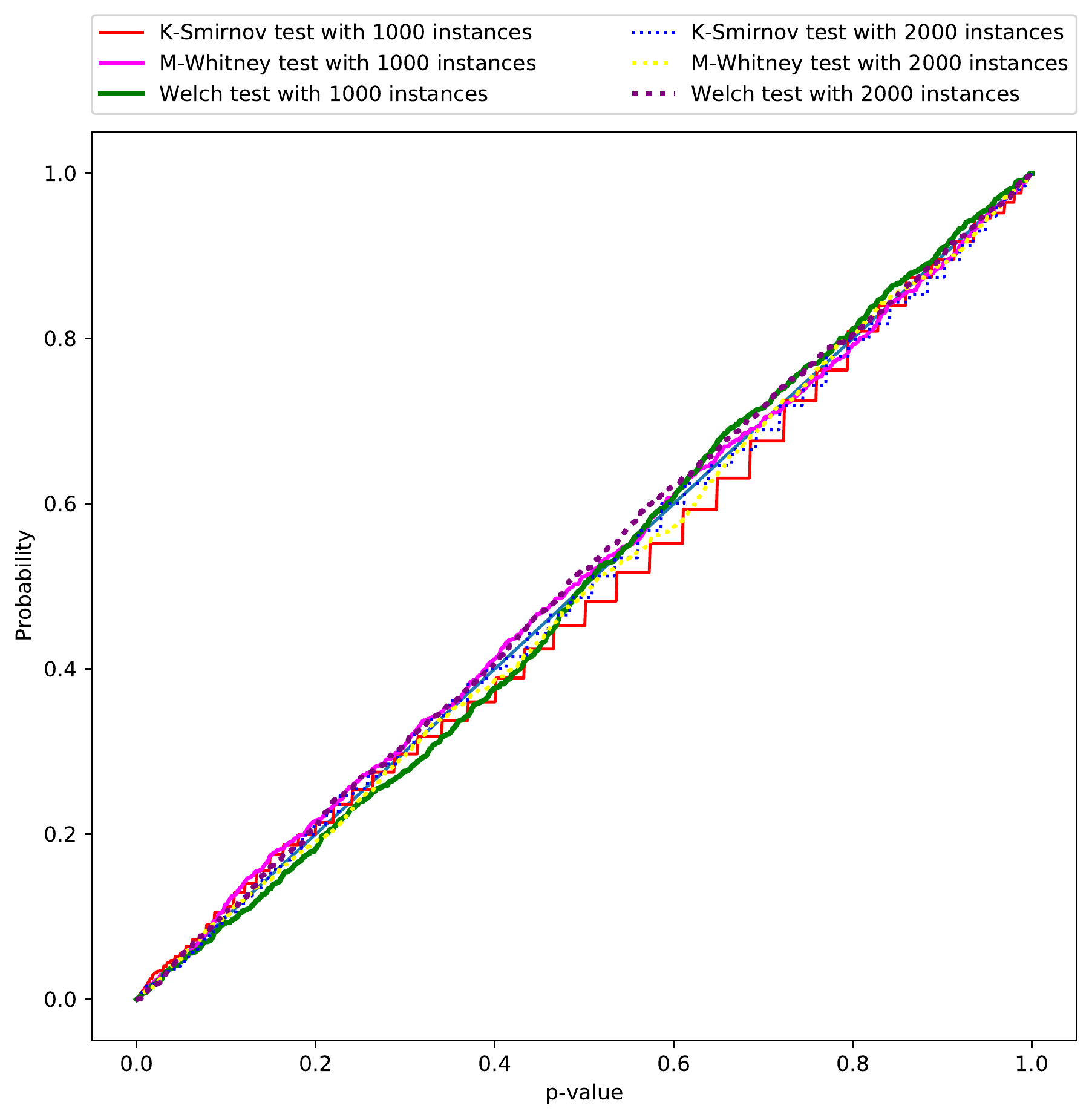}
\centering
\caption{Null hypothesis p-value distribution.}
\label{fig:sstudy_null}
\end{figure}

\begin{figure}[hbtp]
\includegraphics[width=0.75\columnwidth]{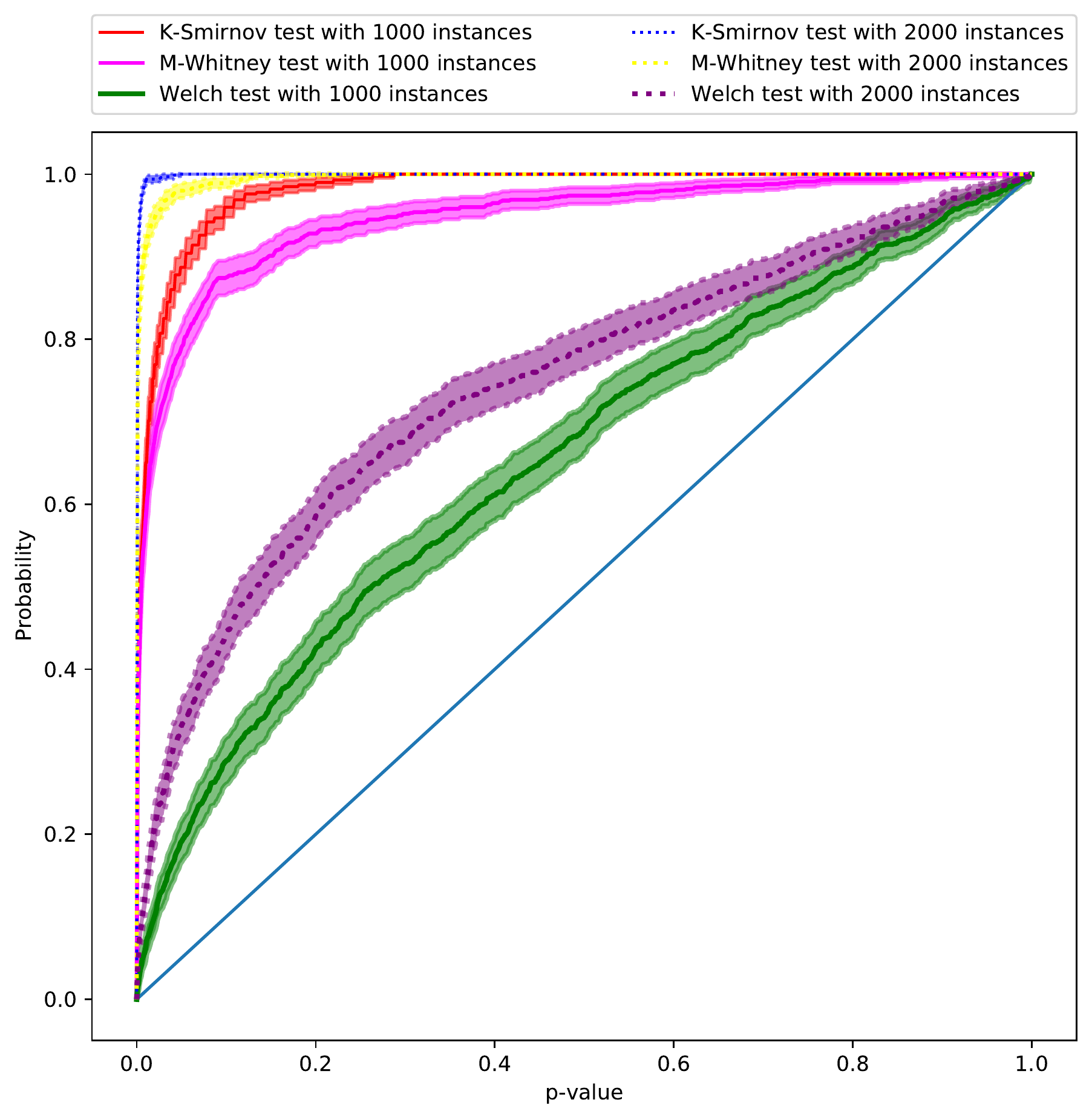}
\centering
\caption{Test power.}
\label{fig:sstudy_power}
\end{figure}

In Figure \ref{fig:sstudy_null}, we present the empirical cumulative distribution of the p-values under the null hypothesis while in Figure \ref{fig:sstudy_power}, we present empirical cumulative distribution of the p-values under the alternative hypothesis, which can also be interpreted as the test power. We also present the confidence bandwidth of two times the standard error (approximately 95\% asymptotically). We also present the results in Tables \ref{tab:sstudy_null} and \ref{tab:sstudy_power}.

\begin{table}[htbp]
 \footnotesize
 \ttfamily
 \centering
 \caption{Test type I error rates.}
 
\begin{tabular}{lrlll}
\toprule
    method &  n instances &     avg p-value &   Error ($\alpha$=1\%) &   Error ($\alpha$=5\%) \\
\midrule
 K-Smirnov &         1000 &  0.511 (0.009) &  0.016 (0.004) &  0.056 (0.007) \\
 K-Smirnov &         2000 &  0.505 (0.009) &  0.015 (0.004) &  0.044 (0.006) \\
 M-Whitney &         1000 &  0.494 (0.009) &  0.012 (0.003) &  0.049 (0.007) \\
 M-Whitney &         2000 &  0.506 (0.009) &  0.010 (0.003) &  0.050 (0.007) \\
     Welch &         1000 &  0.502 (0.009) &  0.008 (0.003) &  0.047 (0.007) \\
     Welch &         2000 &  0.490 (0.009) &  0.006 (0.002) &  0.055 (0.007) \\
\bottomrule
\end{tabular}

 \label{tab:sstudy_null}
\end{table}

\begin{table}[htbp]
 \footnotesize
 \ttfamily
 \centering
 \caption{Test power.}
 
\begin{tabular}{lrlll}
\toprule
    method &  n instances &     avg p-value &   Power ($\alpha$=1\%) &   Power ($\alpha$=5\%) \\
\midrule
 K-Smirnov &         1000 &  0.019 (0.001) &  0.651 (0.015) &  0.887 (0.010) \\
 K-Smirnov &         2000 &  0.001 (0.000) &  0.991 (0.003) &  1.000 (0.000) \\
 M-Whitney &         1000 &  0.052 (0.004) &  0.580 (0.016) &  0.791 (0.013) \\
 M-Whitney &         2000 &  0.004 (0.001) &  0.910 (0.009) &  0.980 (0.004) \\
     Welch &         1000 &  0.348 (0.009) &  0.062 (0.008) &  0.190 (0.012) \\
     Welch &         2000 &  0.256 (0.009) &  0.127 (0.011) &  0.333 (0.015) \\
\bottomrule
\end{tabular}
 \label{tab:sstudy_power}
\end{table}

As can be seemed from the results, all tests are well behaved in terms of being uniform under the null hypothesis as well as having test power that increases with number of instances, with
the Kolmogorov-Smirnov test outperforming both tests and the Mann-Whitney test outperforming the Welch's t-test.

The source code for this experiment is available to download in the package \textit{examples/hypothesis\_testing} folder at \url{https://gitlab.com/marcoinacio/sstudy/-/tree/master/examples/hypothesis_testing}.

\subsection{Neural networks and non-deterministic estimators}
For neural networks and other non-deterministic estimators, in general, we also randomize the initialization parameters of the estimator\footnote{e.g.: for neural networks, using random Xavier \citep{xavier} or Kaimining \citep{kaimaning} initializations.}

In this case, suppose that the method becomes deterministic given a vector of parameters\footnote{For neural networks, that would be the initial value of its neurons.} $\beta$, we would then be estimating:
\begin{align*}
    E_{D \in \mathbb{D}_P}\left[\mbox{loss}\left(M_{\{k,\beta\}}, D\right)\right]
\end{align*}

Therefore, the non-determinism of the method would be ``averaged out'' after a large number of simulations and the same conclusions would follow as was previously done.

In order to illustrate this, we present an simulation study comparing the performance of neural networks given distinct number of hidden layers, with or without dropout \citep{dropout} and on training and test datasets.

We work with the following true distribution:
\begin{align*}
& X_{i,1} \sim \mbox{Gaussian}(0, 1) \\
& X_{i,2} \sim \mbox{Gaussian}(0, 1) \\
& Y_i \sim \mbox{cos}(X_{i,1}) + \mbox{sin}(X_{i,2}) + e_i \\
& e_i \sim \mbox{Gaussian}(0, 1)
\end{align*}
With each train dataset composed of $((X_1, Y_1), (X_2, Y_2), ..., (X_{1000}, Y_{1000}))$ and each test dataset composed of $((X_{1001}, Y_{1001}, (X_{1002}, Y_{1002}), ..., (X_{2000} ,Y_{2000}))$, both with instances sampled i.i.d. We work with standard dense neural networks with 2 hidden layers of the same size, ELU activations \citep{elu} and batch normalization \citep{batch-normalization}, moreover we use Pytorch \citep{NEURIPS2019_9015} as neural networks framework with \textit{nnlocallinear} Python package \citep{1910.05206} on top it.

\begin{figure}[hbtp]
\includegraphics[width=0.75\columnwidth]{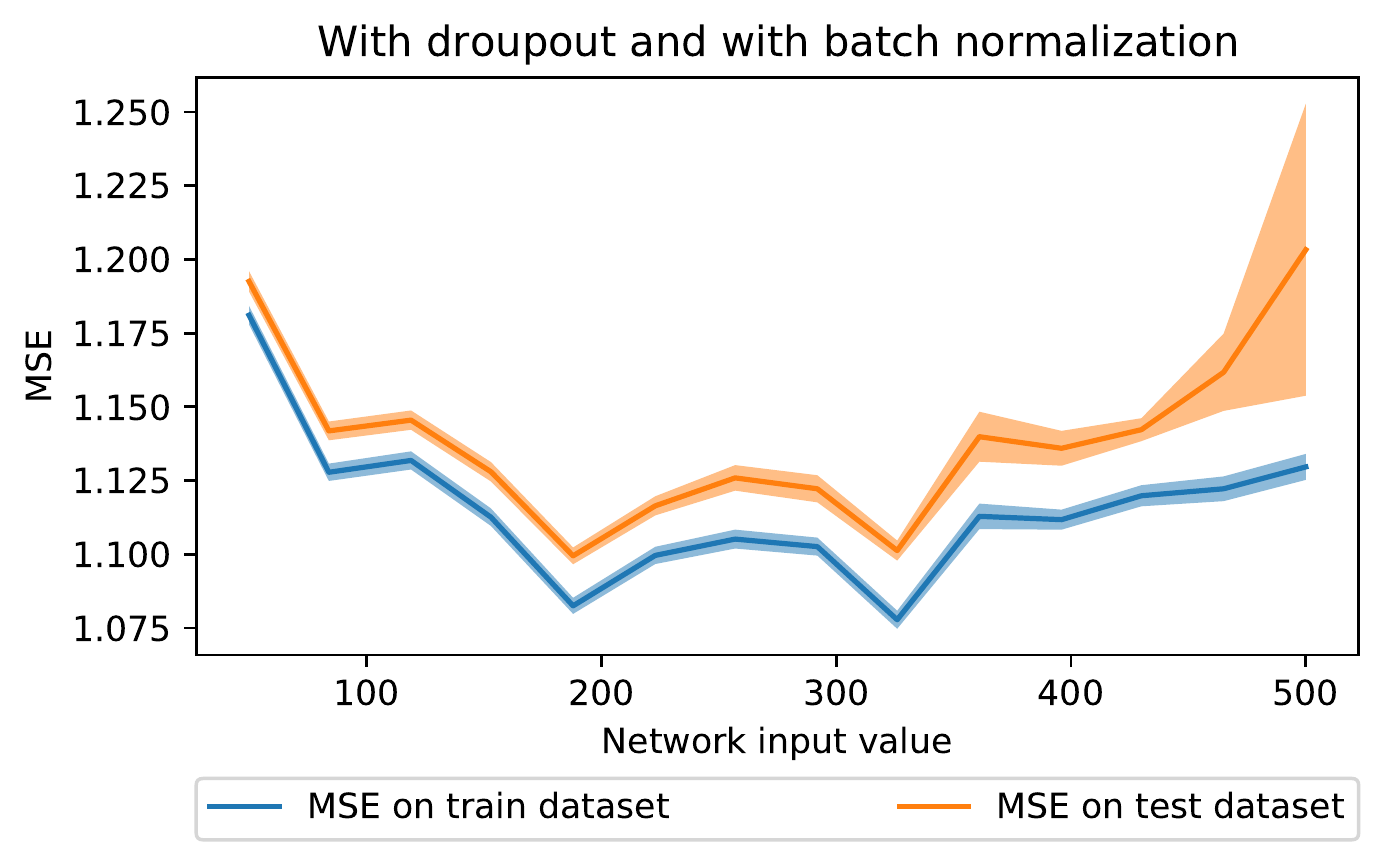}
\centering
\caption{MSE for distinct hidden sizes, with dropout and with batch normalization}
\label{fig:with_dropout_with_batch_normalization}
\end{figure}

\begin{figure}[hbtp]
\includegraphics[width=0.75\columnwidth]{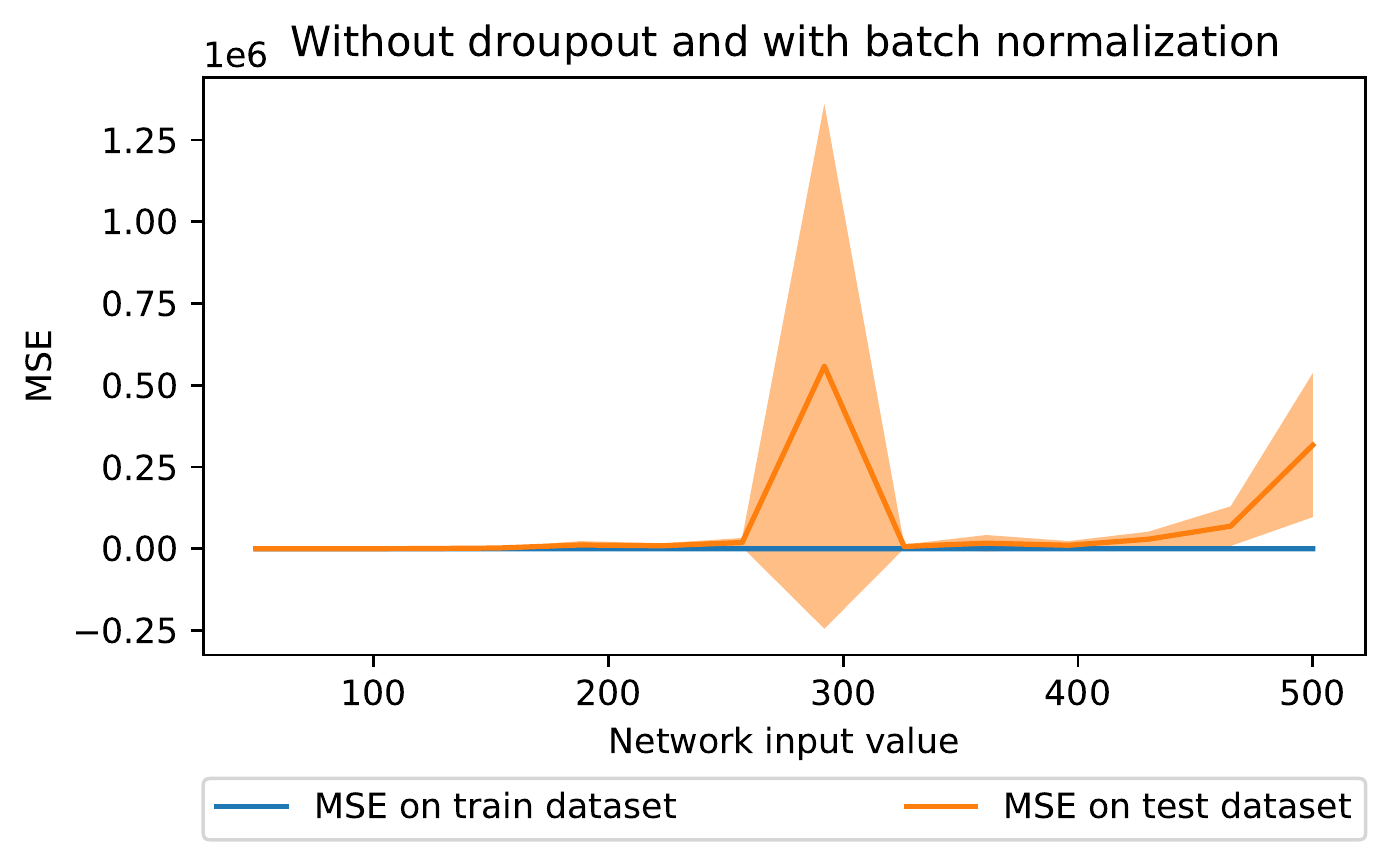}
\centering
\caption{MSE for distinct hidden sizes, without dropout and with batch normalization}
\label{fig:without_dropout_with_batch_normalization}
\end{figure}

\begin{figure}[hbtp]
\includegraphics[width=0.75\columnwidth]{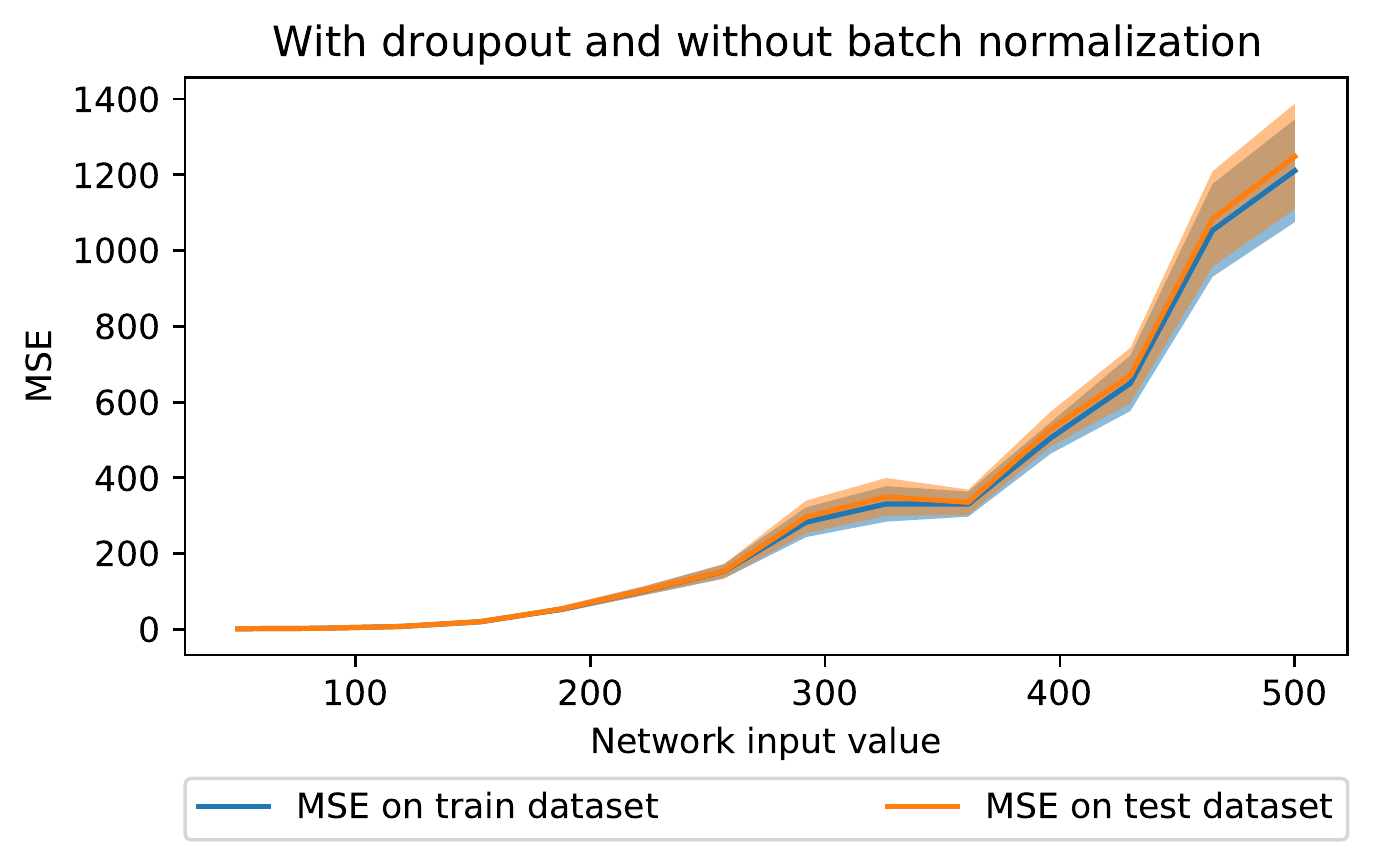}
\centering
\caption{MSE for distinct hidden sizes, with dropout and without batch normalization}
\label{fig:with_dropout_without_batch_normalization}
\end{figure}

\begin{figure}[hbtp]
\includegraphics[width=0.75\columnwidth]{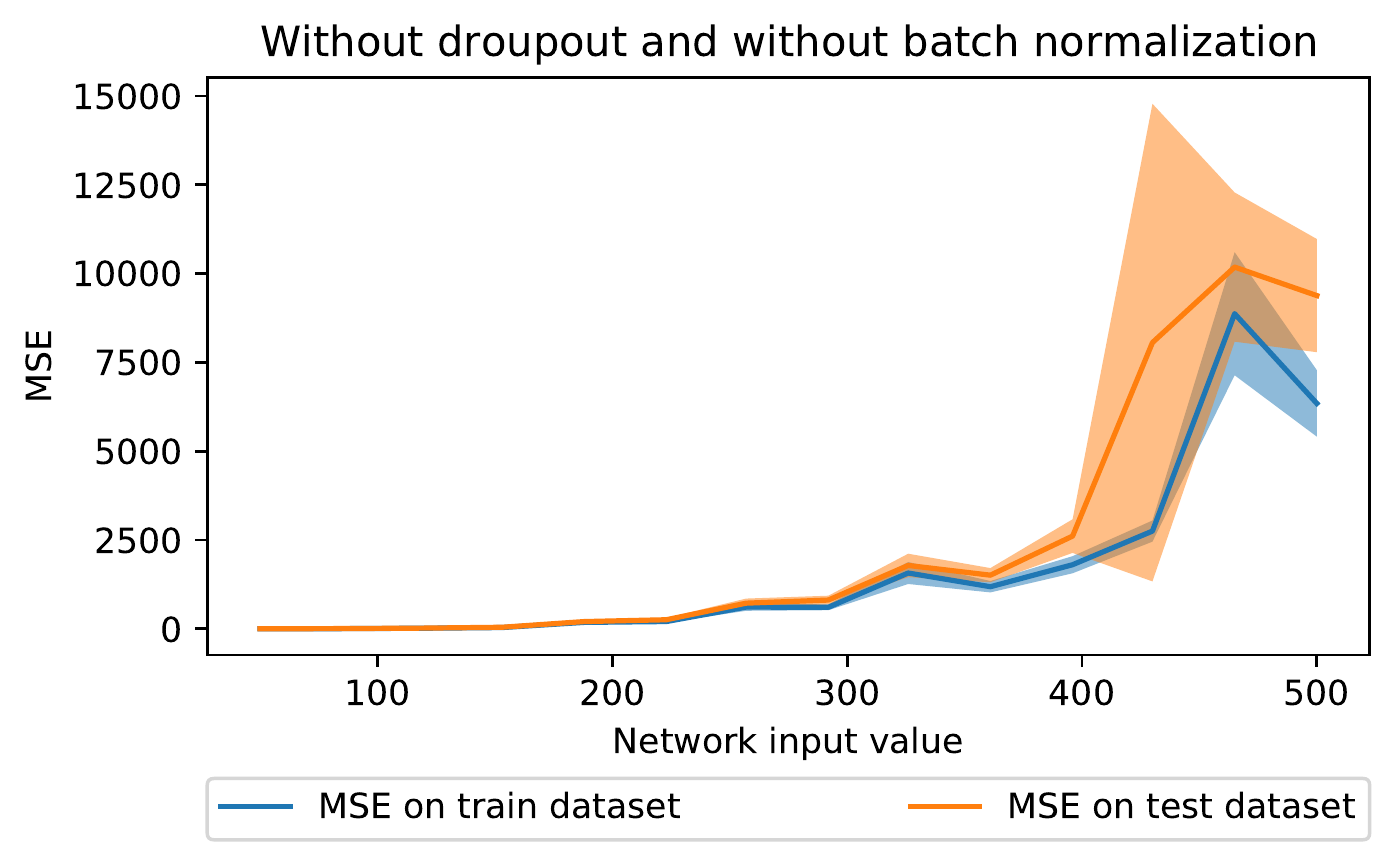}
\centering
\caption{MSE for distinct hidden sizes, without dropout and without batch normalization}
\label{fig:without_dropout_without_batch_normalization}
\end{figure}

In Figures \ref{fig:with_dropout_with_batch_normalization}, \ref{fig:without_dropout_with_batch_normalization}, \ref{fig:with_dropout_without_batch_normalization} and \ref{fig:without_dropout_without_batch_normalization} we present the results of the experiment when with and without dropout and/or batch normalization. When using dropout, we set a 0.5 dropout rate.



The source code for this experiment is available to download in the package \textit{examples/neural\_networks} folder at \url{https://gitlab.com/marcoinacio/sstudy/-/tree/master/examples/neural_networks}.

\subsection{Density estimation and Bayesian inference}
Another interesting instance for using simulation studies is on density estimation. Moreover, Bayesian estimators in general can also have their frequentist properties evaluated using simulation studies \citep{rubin1984bayesianly}.

To illustrate both points, we compare the performance of a npcompare\citep{deAlmeidaIncio2018}: a Bayesian density estimator against the kernel density estimator \citep{kde1, kde2, kde3} with bandwidth hyper parameter chosen by data splitting.

We work with a mixture of gammas as true distribution for generating the dataset $(Y_1, Y_2, ..., Y_n)$:
\begin{align*}
& X_{i,1} \sim \mbox{Beta}(1.3, 1.3) \\
& X_{i,2} \sim \mbox{Beta}(1.1, 3.0) \\
& X_{i,2} \sim \mbox{Beta}(5.0, 1.0) \\
& X_{i,2} \sim \mbox{Beta}(1.5, 4.0) \\
& P(Y_i = X_{i,1}) = 0.2 \\
& P(Y_i = X_{i,2}) = 0.25 \\
& P(Y_i = X_{i,3}) = 0.35 \\
& P(Y_i = X_{i,4}) = 0.2 \\
\end{align*}
Moreover, we use the integrated squared loss as loss function:
\begin{align*}
    \int_0^1 (f(x) - \hat{f}(x))^2 \mathrm{d}x
\end{align*}
Note that in this case, there is no test dataset and the loss is evaluated directly against the true distribution.

In Table \ref{tab:density_estimation}, we present the results of the experiment

\begin{table}[htbp]
 \ttfamily
 \centering
 \caption{Results for a density estimation experiment.}
 
\begin{tabular}{lllr}
\toprule
number & \multirow{2}{*}{method}  & \multirow{2}{*}{loss} &  number \\
no instances & & & simul \\
\midrule
\multirow{2}{*}{100} & kde &  0.107 (0.003) &             500 \\
    & npcompare &  0.038 (0.005) &              30 \\
\cline{1-4}
\multirow{2}{*}{200} & kde &  0.068 (0.002) &             500 \\
    & npcompare &  0.021 (0.002) &              30 \\
\bottomrule
\end{tabular}
 \label{tab:density_estimation}
\end{table}

As can be seemed, the \textit{npcompare} method outperfomed the \textit{kde} for with both 100 and 200 instances. Note that we used a lower number of simulations for the \textit{npcompare} method to its higher computational time, however, this was enough to notice the superiority of the method (for this true distribution) given the calculated standard error.

The source code for this experiment is available to download in the package \textit{examples/density\_estimation} folder at \url{https://gitlab.com/marcoinacio/sstudy/-/tree/master/examples/density_estimation}.

\section{Conclusion}
\label{sec:article_conclusion}

In this short \paper, we have presented a Python package called \textit{sstudy}, designed to simplify the preparation of simulation studies; we presented its basic features, usage examples and references to the its documentation. Moreover, we also presented a short statistical description of the simulation study procedure as well as usage examples.

\section*{Acknowledgments}
Marco In\'{a}cio is grateful for the financial support of CAPES (this study was financed in part by the Coordena\c{c}\~{a}o de Aperfei\c{c}oamento de Pessoal de N\'{\i}vel Superior - Brasil (CAPES) - Finance Code 001), of the Erasmus Plus programme and of the BME-Artificial Intelligence FIKP grant of Ministry of Human Resources (BME FIKP-MI/SC).

\FloatBarrier

\printbibliography

\end{document}